\documentclass[12pt]{article}

\title{Balancing Security and Liquidity: A Time-Weighted Snapshot Framework for DAO Governance Voting} 
\usepackage{orcidlink}
\author{Zayn Wang\orcidlink{0009-0006-4133-8181}, Frank Pu, Vinci Cheung, Robert Hao \\
\\
New York University}

\date{April 2025}

\usepackage{amssymb}
\usepackage{amsmath}

\begin{document}
\maketitle

\begin{abstract}
    As new project upgrading the blockchain industry, novel forms of attack challenges developers to rethink about the design of their innovations. In the growth stage of the development, Decentralized Autonomous Organizations (DAO) introduces different approaches in managing fund through voting in governance tokens. However, relying on tokens as a weight for voting introduces opportunities for hackers to manipulate voting results through flash loan, allowing malicious proposals -  fund withdrawal from DAO to hacker's wallet - to execute through the smart contract. In this research, we learned different defense mechanism against the flash loan attack, and their weakness in accessibility that compromise the security of different blockchain projects. Based on our observation, we propose a new defensing structure and apply it with cases.
\end{abstract}

\section{Introduction and Related Works}


This section provides a systematic review of decentralized autonomous organizations (DAOs), governance mechanisms, flash loans, and their associated security risks. By analyzing these components and their interplay, we contextualize the vulnerabilities exploited in flash loan attacks and evaluate existing defense strategies.

\subsection{Decentralized Autonomous Organizations (DAOs)}

Decentralized autonomous organizations (DAOs) are blockchain-based entities run under smart contracts and community voting, therefore facilitating decentralized decision-making.\cite{arxiv2024}\cite{frontiers2025} There are several important characteristics that define DAO. First, Smart contracts, which will run independently without any intermediary, capture DAO rules and operations. For instance, Uniswap has a fully automated fee distribution mechanism through smart contract architecture. Second, usually token based, DAO membership gives governance powers commensurate with token ownership. For example, Marker DAO gives MKR to its participants for their voting authority. Third, Dao operates transparently, as all proposals and votes are recorded on-chain for public verification. Despite its creative design, DAO suffers from numerous attacks since its inception. The 2016 DAO hack, which exploited reentrancy vulnerabilities to drain 60 million in funds, revealed critical weaknesses in early DAO formations.\cite{acm2024} Modern DAOs such as Aragon and DAOstack have embraced modular governance models to improve security and adaptability, but their reliance on token-weighted voting establishes new attack channels for hostile parties.\cite{ssrn2025}

\subsection{Governance in DAO Ecosystems}

DAO governance systems control the proposed, discussed, and passed modifications in protocols.\cite{ssrn2025} Although these systems are fundamental for DAOs' operation, they also provide weaknesses that attackers could find use for. Token-weighted voting—where holders of governance tokens (such as COMP or UNI) vote on proposals commensurate with their protocol stake—is a common governance technique. Although this system encourages democratic decision-making, it is susceptible to flash loan governance attacks—where attackers borrow tokens momentarily to acquire voting authority and forward destructive proposals.\cite{acm2024} To prevent such attacks, Compound's Governor Bravo proposal introduced a two-day await for votes on recently acquired tokens. This system has successfully dropped governance takeover rates. Another governance mechanism is delegated voting, where token holders delegate their voting power to trusted representatives or experts. Although this strategy increases the efficiency of decision-making, it exposes hazards should delegates' wallets be hacked or if they take bribes from hostile agents.\cite{ssrn2025} For instance, attackers used weaknesses in delegation contracts in the 2023 Frax Finance incident to control governance outcomes.

It is important to note that DAO is a representation of DeFi. Its governance structure, unlike the traditional stock based centralized organizations that have an A/B type stock and represent a centralized control. Most DAOs rely on token weighted based voting. The closest parallel to A/B type stock is multi-sig committees where a small group holds veto or execution power. This is not a common practice.

\subsection{Flash Loan: Mechanism and Applications}

Flash loans are uncollateralized loans that must be repaid within a single transaction block. They are a novel financial instrument unique to blockchain technology and are widely used across decentralized finance (DeFi) platforms such as Aave and dYdX.\cite{hacken2023}\cite{nadcab2024} Flash loans operate based on atomic execution principles; if any step in the transaction sequence (borrow, execute operations, repay) fails, the entire transaction is reverted on-chain. Flash loans offer several legitimate use cases that contribute to DeFi innovation. They enable arbitrage opportunities by allowing traders to exploit price differences across exchanges without risking their own capital. Flash loans also facilitate collateral swapping for users seeking more favorable collateralization terms and debt refinancing for users consolidating loans across protocols. However, the same characteristics that make flash loans attractive for legitimate purposes also enable malicious exploitation. In 2023 alone, platforms like Aave processed \$12.3 billion in flash loan volume—some of which was used for attacks targeting vulnerable protocols.

\subsection{Flash Loan Attacks: Pattern and Impacts}

Flash loan attacks exploit the atomicity of transactions and vulnerabilities in smart contracts or governance mechanisms to execute malicious operations without requiring upfront capital. Price oracle manipulation is a typical attack vector whereby attackers artificially inflate or deflate asset prices within low-liquidity pools or faulty oracle systems using flash loans.\cite{acm2024} This strategy was used in the Mango Markets hack (2022), in which attackers manipulated pricing to borrow overvalued assets and drained \$117 million from the protocol. Another vector of attack involves governance takeovers\cite{ssrn2025}, where attackers borrow governance tokens using flash loans to pass fraudulent proposals that benefit them at the expense of the protocol or its users. An example of such is the 2024 XToken case where attackers borrowed tokens to approve a proposal to directly transfer funds to their wallet. The economic impact of flash loans was devastating, between 2020 to 2024, there are over billions of dollars stolen from flash loan attacks solely. Beyond the financial losses, these attacks also erode users' confidence in DAO as a secure organization to hold their funds. \cite{acm2024}\cite{ssrn2025}

\subsection{Defense Mechanism Against Flash Loan Attacks}

To address the growing threat of flash loan attacks, researchers and developers have proposed various defense mechanisms targeting specific vulnerabilities.\cite{acm2024} Technical protections include reentrancy guards that prevent recursive function calls during transaction execution. Chainlink utilizes this strategy to mitigate basic flash loan attacks. Governance protections include time-delayed voting rights for new acquired tokens. This prevents immediate leveraging of borrowed tokens to take over the protocol. Temporal control mechanisms represent another line of flash loan defense mechanisms by disrupting the atomic transaction sequence. Static holding period imposed a fixed delay on tokens transfer or usage after it is being acquired. Dynamic holding periods offer an adaptive alternative by adjusting the lock time based on the total value of the tokens.\cite{arxiv2024}\cite{acm2024}\cite{ssrn2025}
\subsection{Synthesis}

Although DAOs and flash loans have driven considerable amounts of innovation in decentralized finance ecosystems, their integration generates systematic hazards that call for strong mitigating techniques. While current solutions solve specific weaknesses, they do not offer complete defense against multi-vector attacks that concurrently target smart contract logic faults and governance systems. By suggesting adaptive temporal restrictions coordinated with snap-shot based monitoring systems to improve resistance against changing attack strategies while maintaining operational efficiency across DeFi protocols, our work closes this gap.

\section{Defense Structures}

\subsection{Recent solutions}

In response to the increasing sophistication and frequency of flash loan attacks on DAOs, the blockchain community has developed a diverse set of countermeasures aimed at reinforcing governance integrity. This section provides a broad review of both popular and emerging solutions, while offering a deep dive into their mechanisms, strengths, and limitations in the context of DAO vulnerabilities. These solutions are not only technical in nature but also institutional, targeting the governance-layer weaknesses that flash loans often exploit, namely the ease of acquiring large voting power instantly and executing malicious proposals within a short timeframe. 

\subsubsection{Snapshot-Based Voting}

Snapshot-based voting is a preventive mechanism where token balances are recorded at a specific block prior to the commencement of a governance vote. Voting power is determined by this snapshot, preventing tokens acquired afterward from influencing the outcome, such as those obtained via flash loans. While traditionally a single snapshot is taken before voting begins, a more robust approach involves taking snapshots continuously or at regular intervals, enhancing the system’s ability to detect and neutralize flash loan-based attacks. Under this model, voting power is effectively based on consistent historical presence rather than transient ownership, thereby discouraging opportunistic governance manipulation. This dynamic snapshot strategy primarily targets tokens with short holding durations, significantly reducing the impact of flash-loaned assets.\\
A key advantage of this approach is its simplicity and efficiency: off-chain snapshots, commonly used by platforms like Uniswap and Aave, reduce gas fees while maintaining verifiability against on-chain data. However, its static nature can be a drawback. Newcomers who acquire governance tokens after the snapshot are excluded from participation, potentially discouraging new user engagement. Additionally, reliance on off-chain infrastructure introduces centralization risks if trusted parties are involved in snapshot creation. Overall, snapshot-based voting is a foundational layer of defense, particularly effective when coupled with continuous tracking and time-aware governance logic.

\subsubsection{Time-weighted voting}

Time-weighted voting enhances governance integrity by tying voting power not only to the number of tokens held but also to the length of time they’ve been held. This strategy neutralizes flash loan attacks by assigning minimal or zero weight to tokens with zero holding duration, effectively voiding the influence of freshly borrowed tokens. \\
Time-weighted mechanisms also encourage long-term community participation and deter governance volatility, as only users with sustained commitment to the project can meaningfully influence decisions. However, the implementation of such systems is technically complex. It requires precise tracking of token age, increasing smart contract surface area, gas usage, and potential vulnerability to bugs. Moreover, it can alienate new investors who initially have limited voting power, and entrenched holders may accumulate excessive influence even if their interests no longer align with the DAO. In early-stage DAOs, where no participant has long holding periods yet, time-weighted voting offers limited protection. Therefore, while the strategy is highly effective in established ecosystems, its use requires careful design and community onboarding strategies.

\subsubsection{Token holding period}
Token holding period imposes a delay between acquiring tokens and being eligible to vote. Governance tokens must be held or locked for a minimum duration before conferring any voting power, ensuring that temporary holdings cannot be used for manipulative purposes, such as those obtained via flash loans. This approach directly restricts market timing attacks and strengthens the credibility of the governance process by aligning it with token holder commitment. Examples like Curve’s veCRV and Balancer’s veBAL systems have popularized vote-escrowed tokenomics that scale voting power based on lock-up durations. \\
Token holding periods operate through smart contract-enforced time locks integrated at the protocol level. With transaction interceptions, smart contracts monitor token transfers, flagging assets obtained through flash loan mechanisms. Flagged tokens are automatically placed in escrows for a protocol-defined duration (e.g., 1 hour to 7 days). During the holding period, locked tokens cannot be used as collateral, participate in governance votes, nor be transferred to external addresses. Assets meet releasing conditions will unlock automatically after the elapsed time or upon satisfying secondary verification checks. \\
\begin{figure}[!h]
      \centering
      \includegraphics[width=0.5\linewidth]{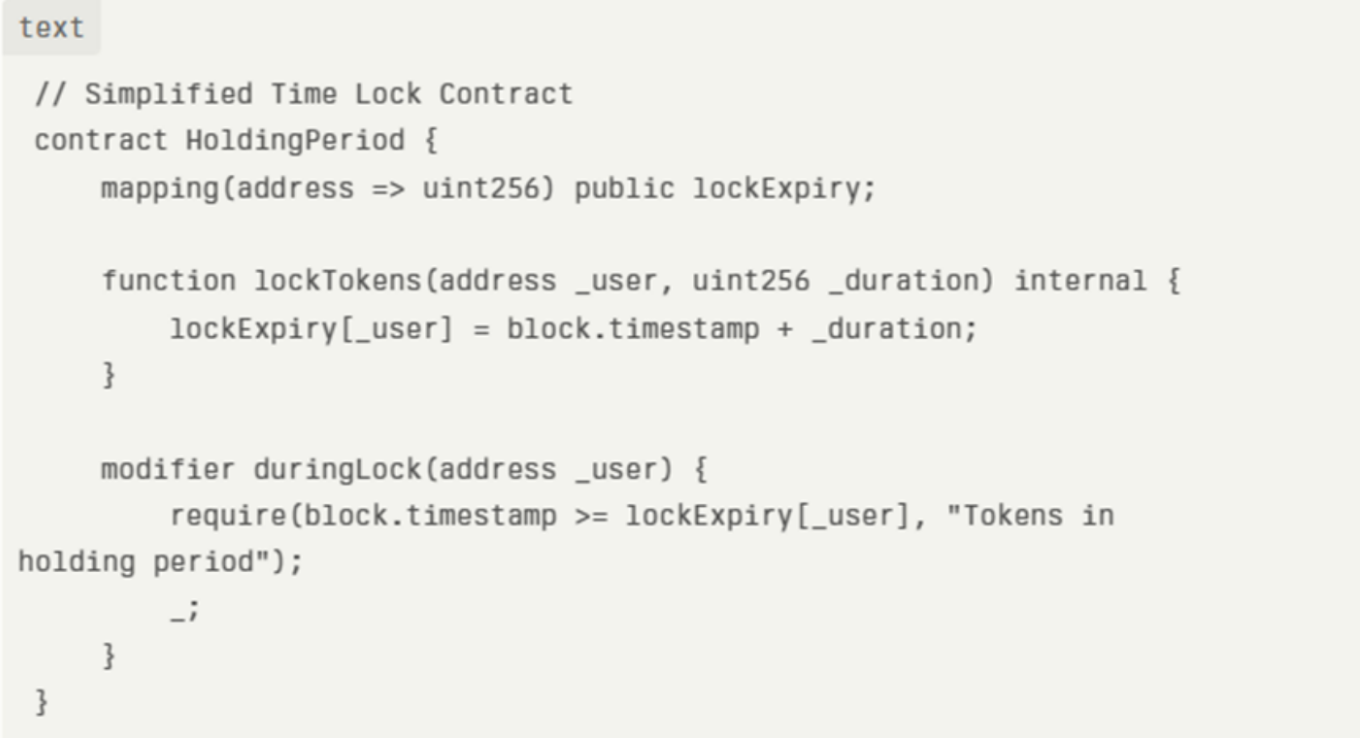}
      \caption{Holding period code example}
      \label{fig1}
\end{figure} \\
The upside lies in discouraging rapid speculation and enhancing resistance to governance capture. Yet, these measures also come with trade-offs. Liquidity is reduced as users must forgo short-term flexibility, and centralization risks arise as long-term holders disproportionately gain influence. Additionally, implementation requires governance-compatible staking and lock-up infrastructure, which increases operational complexity. This solution suits DAOs that seek to combine stability with gradual participation, especially in environments where token distribution is expected to remain consistent over time.

\subsubsection{Complementary and reactive defenses}
In addition to the three primary solutions, several complementary approaches provide secondary lines of defense. \textbf{Flash loan detection} mechanisms, such as those deployed by Forta and FlashGuard, operate by analyzing blockchain data in real-time to flag patterns indicative of flash loan behavior. While useful for rapid alerts and post-factum reaction, these systems are primarily reactive and can suffer from false positives, high computational requirements, and delayed intervention. \textbf{Time-locked proposals} introduce a delay between proposal approval and execution, granting the community time to counteract any malicious governance actions. However, this slows down legitimate governance processes and may introduce implementation bugs. Another adaptive strategy is \textbf{dynamic quorum and threshold adjustment}, where the participation and approval requirements change based on proposal importance or controversy. This deters low-turnout manipulation but adds complexity to governance rules and risks confusing stakeholders. Finally, \textbf{multi-signature wallets} introduce execution-layer security by requiring multiple keyholders to approve transactions, is a reliable failsafe for treasury protection. While not strictly a voting safeguard, multi-sig arrangements ensure that even if a flash loan attack succeeds in governance, malicious fund withdrawals can be halted. The coordination burden and key management overhead, however, make them more appropriate for backend fund control than for user-facing governance.

\subsection{Our solution}

Inspired by attacks through oracle manipulation on cream finance in Appendix \ref{cream} and the liquidity problem, our solution involved two methods: snapshot-based voting and time-weighted voting, trying to solve the problems of indirect flash loan attack and liquidity. Here is how our solution goes:
\begin{itemize}
  \item Preparing for a vote: Taking a snapshot $m$ $S_m$ of each block $n$, where snapshot $S_m = d:t_d^m$ or $S_m(d) =t_d^m$ is a dictionary that takes the address $d_m$ (or its hash value) as the key and returns the amount of tokens of the address in that snapshot $t_d^m$ as a value.
  \item Determining the Dynamic weight function: the weight function is a function for a specific address (or its hash value) that maps to their tokens in each block which is a vector $t_d = (t_d^0, t_d^1, \ldots, t_d^m)$ to a real number. The weights could be presented as a vector $w = (w_0, w_1, \ldots, w_m)$, which is designed or adjusted by the community or a central third party; then the weight function could be represented as $f(t_d) = w \cdot t_d$. More about dynamicity: since the only factor that affects the weight function is the weight vector $w$, and the weight vector is easy to understand and visualized by plotting with statistical methods. It is easy to be reviewed by the community and change the vector by voting/the centralizing third party.
  \item File a vote: the proposer files a governance proposal and each address comes into voting.
  \item Weight functions as voting power. For each address, the voting power $p_d = f(t_d)$ is determined by the weight function and the tokens the address holds in each snapshot.
  \item (time-locked period could also be introduced for community to review)
  \item Approve/Disapprove for voting: Sum the weighted votes for approval of the proposal $v_a = \sum_{d \in D_a}p_d$, comparing with whom the disapproved $v_d = \sum_{d \in D_d}p_d$ then decides whether or not a proposal is approved. Then, the result is also a function $V: \mathbb{R} \times \mathbb{R} \to \{A, D\}$, for example, $V(v_a, v_d) = \begin{cases} A & v_a \geq \frac{p}{1 - p}v_d \\ D & \text{else} \end{cases}$ where in range $A$ stands for approve a proposal and $D$ stands for disapproval, and $v_a \geq \frac{p}{1 - p}v_d$ means $v_a$ has to at least be in $p$ proportion to approve a proposal.
\end{itemize}
To sum up, the solution for our model is $V(v_a, v_d)$, where $v_a = \sum_{d \in D_a}p_d$ and $v_d = \sum_{d \in D_d}p_d$,  $p_d = f(t_d)$, $f(t_d) = w \cdot t_d$ where  $t_d = (t_d^0, t_d^1, \ldots, t_d^m)$ and $w = (w_0, w_1, \ldots, w_m)$.

\section{Experiments and Cases Study}

\subsection{Recent cases}

\begin{itemize}
  \item Case 1: Beanstalk DAO - Flash Loan Governance Exploit  
  \subitem Beanstalk Attack
   \begin{figure}[!h]
      \centering
      \includegraphics[width=0.5\linewidth]{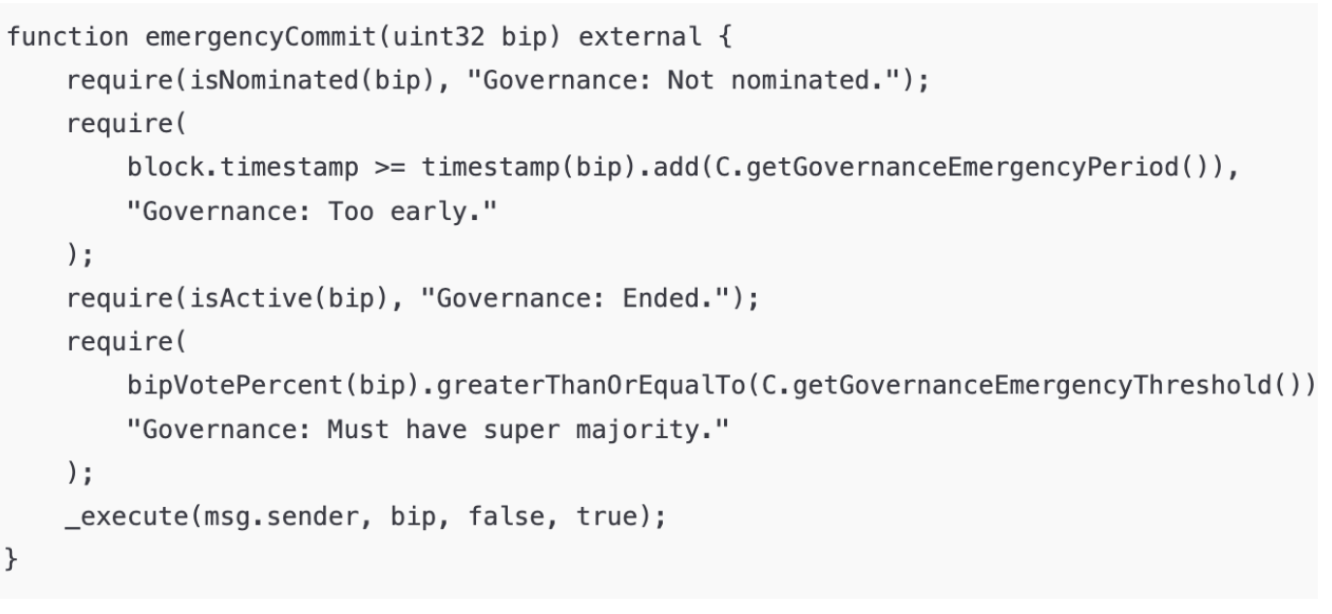}
      \caption{Beanstalk emergencyCommit function}
      \label{fig2}
  \end{figure}
On April 17, 2022, the Beanstalk DAO, or a credit-based stablecoin protocol, suffered a devastating governance attack that exploited vulnerabilities in its snapshot-based voting system. At the time, Beanstalk relied on off-chain snapshot voting with on-chain execution, using the emergencyCommit function to rapidly deploy patches or adjust protocol parameters during periods of volatility. Figure 1 displays the emergencyCommit function, executed under which a 24-hour wait period (line 4) and 66\% of Liquitidy Pool Tokens (line 9) are met.\cite{verichains2022beanstalk}  The attacker capitalized on the lack of a delay between token acquisition and voting influence: snapshot voting in Beanstalk granted immediate governance power to newly acquired tokens, without requiring any holding period. The attacker submitted a malicious Bean Improvement Proposal (BIP-18), referencing a precomputed smart contract address created using Ethereum’s CREATE2 opcode. Since the contract was not deployed at the time of proposal submission, its bytecode remained inaccessible on-chain, preventing community review. After the required 24-hour wait period had elapsed, the attacker amassed a flash loan of 1 billion dollars and temporarily amassed 80 percent of Beanstalk’s voting power to execute the emergencyCommit, triggering a single complex transaction containing over 20 operations. These included deploying the malicious contract, voting with the liquidity pool tokens that hold Stalk (governance) power, passing the proposal with a supermajority, and transferring 182 million dollars in funds to their address—all within a single Ethereum transaction. They then repaid the flash loan and kept the remaining stolen assets as profit. In response, Beanstalk implemented a multisig wallet as a human safeguard to prevent code automation, reflecting a changing degree of reliance developers have on the code. \cite{immunefi2022beanstalk} 
  \item Case 2: UPCX - Admin Privilege Exploit
  \begin{figure}
      \centering
      \includegraphics[width=0.5\linewidth]{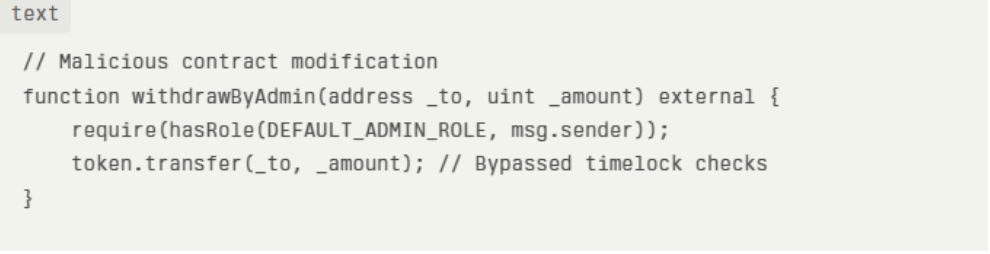}
      \caption{Modified UPCX contract vulnerable to malicious attack}
      \label{fig3}
  \end{figure}
  
  \begin{figure}
      \centering
      \includegraphics[width=0.5\linewidth]{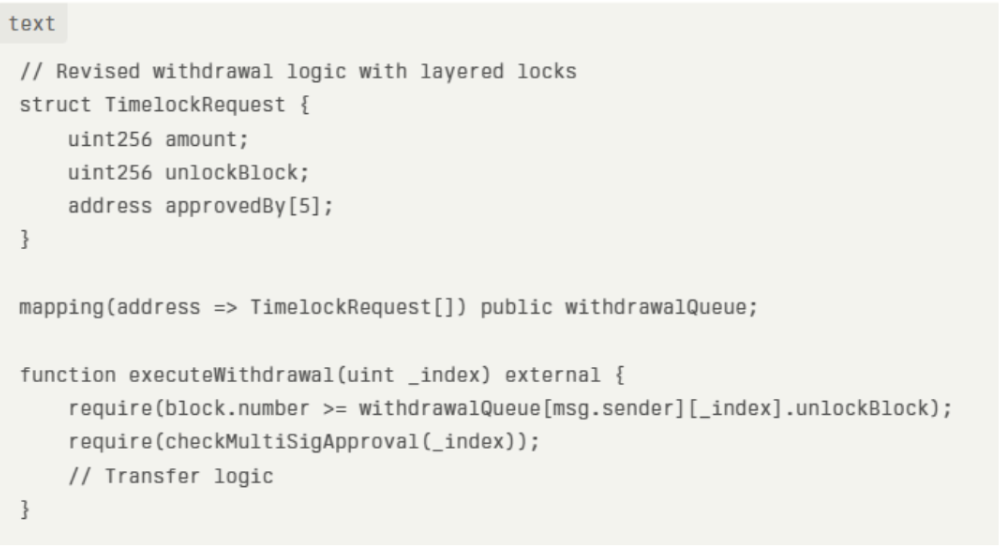}
      \caption{Revised UPCX Withdrawal logic with layered locks}
      \label{fig4}
  \end{figure}
    \subitem On April 1, 2025, the decentralized protocol UPCX suffered a 70 million dollar exploit despite having token holding periods intended to mitigate rapid asset movements. Although UPCX enforced a 24-72 hour holding period for user withdrawals and governance proposals (seven days), administrative functions were not subject to similar temporal controls. The upgrade process lacked multisig verification, cross-contract state checks, and any form of privilege separation. On the other hand, 68 percent of protocol-owned liquidity was concentrated in three administrative wallets, one of which controlled all core contract upgrade permissions and was ultimately compromised. This centralization, combined with incomplete timelock coverage, increases UPCX’s operating risk. \cite{mitrade2025upcx} The attacker compromised an administrative wallet. Having access to upgrade the ProxyAdmin contract, the attacker removes safeguards to create a single point of failure (Figure 2). By modifying the withdrawByAdmin function (figure 3) to bypass the enforced seven-day withdrawal delay, the attacker permits an unauthorized transfer of 18.4 million UPC tokens (2.36 percent of the total supply) in a single transaction. The incident underscores the inadequacy of partial locking mechanisms and the necessity of enforcing holding periods uniformly across all control layers.\cite{cybertecwiz2025upcx}  In response, UPCX implemented an emergency contract freeze and protocol fork. Long-term mitigation strategies included a revised governance framework featuring 5-of-9 multisig authorization for admin actions, block-based rather than timestamp-based locks, and cross-contract state validation. Additionally, the team initiated economic rebalancing by reducing protocol-owned liquidity and introducing dynamic holding periods that adjust with market volatility. \cite{cryptonews2025upcx} 

  \item Case 4: Curve Finance - Time-Weighted Voting Mode
  \begin{figure}
      \centering
      \includegraphics[width=0.5\linewidth]{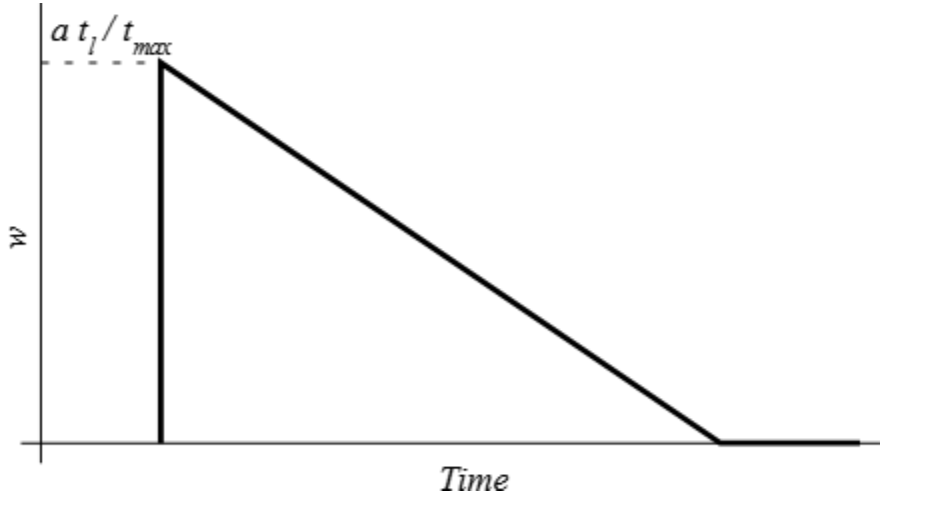}
      \caption{Curve Finance: Visual Representation of Linear decay of Voting Power}
      \label{fig5}
  \end{figure}
    \subitem Curve Finance, a decentralized exchange specializing in stablecoin trading, employs a time-weighted voting mechanism that has proven effective in preventing flash loan governance exploits. \cite{curvefi2025readme} While no successful flash loan attacks have been recorded against Curve’s governance, the protocol offers a robust real-world implementation of time-weighted voting that prioritizes long-term participation over short-term influence.  The mechanism functions through a token locking model: users lock CRV tokens for up to four years, and voting power is calculated as the product of the locked amount and the lock duration. This voting power decays linearly over time as the lock period approaches expiration, incentivizing users to extend or increase their locks to retain influence. \cite{curvefi2025gauges} Governance logic is managed through the VotingEscrow contract, which uses parameters like slope and bias stored in user point history to model each user’s voting power curve. Global changes are tracked in point history, and scheduled adjustments to voting power are recorded via slope changes, ensuring scalable and predictable governance dynamics. However, this model also presents trade-offs. Token locking results in reduced liquidity, potentially deterring users unwilling to immobilize their assets. The system also requires active lock management, adding complexity for participants. Furthermore, long-term locking may lead to centralization, as only a small subset of participants may be willing or able to commit large amounts of CRV over extended durations. \cite{curvefi2025votingescrow} 

\end{itemize}

\subsection{Thematic Analysis: Temporal Misalignment in Governance Design}
Beanstalk and UCPX attacks assumed that holding tokens requires a long-term commitment. However, on-chain realities allow attackers to temporarily acquire tokens, vote, and exit within a single block. Beanstalk reflected the limitations of snapshot-based voting in addressing unforeseen cases beyond just flash loan attacks, allowing attackers to exploit incompatibilities between different temporal defense mechanisms. In UPCX’s case, the flaw was a reliance on time-based assumptions of continuous administrative control, in which changes would only securely occur through authorized actors. This assumption broke down when the off-chain governance failed, and an attacker executed a malicious upgrade on-chain, exploiting the lack of on-chain delays or distributed safeguards. In summary, Beanstalk lacked an effective post-vote timelock, and UPCX lacked time-delayed oversight for admin control. Each protocol assumed a sequential process for critical actions (vote accumulation, collateral growth, or system upgrades) where time is the independent variable that atomic DeFi execution does not inherently guarantee.

\subsection{Comparative Insight: Curve’s Model as a Temporal Defense Architecture}

In contrast, Curve’s governance model is structurally resistant to flash loan governance attacks. By requiring long-term token locking to acquire voting rights, it renders short-term speculative control infeasible and aligns influence with protocol loyalty. The linear decay model provides predictable voting trajectories, while slope changes and historical point tracking accommodate governance decisions by granting time for the community to respond to transient fluctuations. Curve also incorporates execution delays and active lock management, recognizing that even with escrow-based voting, time buffers are essential for security. This design stands in sharp contrast to Beanstalk’s minimally delayed emergencyCommit and UPCX’s instant admin execution. By treating time as a dependent variable controlled by multiple defensive mechanisms, Curve enforces a temporal gap between decision and action, aligning governance with the irreversible nature of blockchain transactions.

\subsection{Governance Design Implications}

A comparative analysis of these two attacks against Curve’s security architecture reveals an inverse relationship between accessibility and security as a result of time dependency. While Curve maximizes governance security through long-term commitment and delayed execution, this comes at the cost of reduced asset liquidity as time becomes respondent. Conversely, the exploited protocols prioritized time as a driver to accessibility, allowing for immediate actions but at the expense of systemic security. This trade-off highlights the challenge in DeFi governance design: balancing automation and resilience against user agency and responsiveness. The presence of human intervention (multisig wallets, protocol suspensions) in these cases further underscores the difficulty of maintaining equilibrium between decentralization and centralized oversight, and the ongoing tension between security and accessibility in DAO architecture.

\subsection{Our solution analysis on the cases}

\subsubsection{Defending against Attacks}

Our solution is able to defend against direct flash loan attacks. As direct flash loan attacks occur, the block including the transition of flash loans has not been established yet. Based on our solution, the snapshots are taken only for the established blocks and would not consider the current flash loan. 

And for indirect flash loan attacks such as oracle manipulation in the Appendix \ref{cream}, our solution could work by adjusting the weight function to have more weight in the older blocks and less weight in the new blocks. For example, a weight vector like $w = (0.5, 0.3, 0.2, 0.1, 0.1)$ would make the most recent snapshot potentially affected by oracle manipulation to be weight only for $10\%$, and the oldest block that has not been affected to as high as $50\%$. In this way, the attackers have to spend more to deliver the attacks, since they have to make the effect of oracle manipulation last longer.

\subsubsection{Liquidity}

Many defenses on flash loan attacks somehow reduce liquidity, since the defenses encourage people to hold their tokens longer. However, in our solution, adjustment of the weight function could restore DAO from the mode of defense against flash loan attacks to encourage liquidity. For example, a weight vector like $w = (0, 0,1, 0.2, 0.2, 0.5)$ taking the most recent tokens into the high amount consideration as $50\%$, and people would be more confident to trade their tokens since they know that after trading their tokens, they would still have a place for governance voting.

\subsubsection{Costs}

For the Memory Complexity, it is $O(m * D)$ since the only time needed to store is the neglectable weight vector and the snapshot $m$ (dictionary); and for time complexity, it is $O(m * D)$ because the only thing to calculate is the weighted function for each address.

\subsubsection{Future Works}

Our solution proposes a structure to defend against direct and indirect flash loan attacks, with consideration of protecting the liquidity of a DAO. In the future, there could be more analysis on how to adjust the weight function, like comparing the adjustment between communities (such as proposing voting for the adjustment for weight function) or decisions under a third party. After the adoption of our solution, there could be more analysis of the weight function through data and experiments, to find out the best weight function that can take both defense and pro-liquidity into consideration. It may also leave a space for a combination to train a machine learning network \cite{wang2022emotion} like CNN \cite{wang2024enhancing} for finding better weight functions in the future.

\section{Acknowledgements}

Special thanks to Professor Hanna Halaburda at Kaufman Management Center, Leonard N. Stern School of Business, New York University for her mentorship, detailed introduction to ETH and DAO in paper \cite{halaburda2025decentralization}, and lecture.

\appendix
\section{Cream Finance - Indirect Oracle Manipulation}\label{cream}
Indirect Flash Loan Exploit via Oracle Manipulation – Cream Finance (2021)

On October 27, 2021, an attacker exploited Cream Finance’s lending protocol by initiating a complex chain of transactions starting with a 500 million dollar worth of DAI flash loan from MakerDAO. The exploit involved leveraging Yearn vaults and Curve pools to manipulate Cream’s price oracle, thereby inflating collateral values and increasing borrowing limits. This enabled the attacker to withdraw unbacked assets, ultimately draining over 130 million dollar from the protocol. Unlike direct governance attacks, this exploit utilized an indirect flash loan vector, which is using the flash loan not to vote or execute proposals, but to manipulate collateral valuations through temporary asset deposits. The exploit was particularly difficult to prevent because the manipulation extended over multiple blocks, exploiting the lack of a mechanism to verify sustained token holdings. Cream Finance promptly suspended all interactions with its Ethereum v1 markets to prevent further exploitation. Even if snapshot-based protections had been in place, they would have failed to detect the temporary nature of the attack, as the snapshot would only capture a transient state. In contrast, a time-weighted mechanism could have mitigated this exploit by requiring sustained asset presence over time. Such mechanisms introduce a form of temporal authentication, ensuring that protocol privileges are reserved for long-term participants and are not accessible through ephemeral, block-level manipulations. \cite{immunebytes2021cream}

\bibliographystyle{plain}
\bibliography{_ref.bib}

\begin{thebibliography}{10}

\bibitem{cryptonews2025upcx}
{Crypto.News}.
\newblock Upcx halts transactions after \$70m security breach.
\newblock \url{https://crypto.news/upcx-halts-transactions-after-70m-security-breach/}, 2025.
\newblock Accessed April 18, 2025.

\bibitem{curvefi2025readme}
{Curve Finance}.
\newblock Curve dao contracts documentation readme.
\newblock \url{https://github.com/curvefi/curve-dao-contracts/blob/master/doc/README.md}, 2025.
\newblock Accessed April 18, 2025.

\bibitem{curvefi2025gauges}
{Curve Finance}.
\newblock Gauge documentation.
\newblock \url{https://github.com/curvefi/curve-dao-contracts/blob/master/doc/gauges.md}, 2025.
\newblock Accessed April 18, 2025.

\bibitem{curvefi2025votingescrow}
{Curve Finance}.
\newblock Votingescrow.vy contract source code.
\newblock \url{https://github.com/curvefi/curve-dao-contracts/blob/master/contracts/VotingEscrow.vy}, 2025.
\newblock Accessed April 18, 2025.

\bibitem{cybertecwiz2025upcx}
{CybertecWiz}.
\newblock Upcx loses \$70 million worth of upc tokens due to security breach.
\newblock \url{https://www.cybertecwiz.com/upcx-losses-70-million-worth-of-upc-tokens-due-to-security-breach/}, 2025.
\newblock Accessed April 18, 2025.

\bibitem{hacken2023}
Hacken.
\newblock Flash loan attacks: Risks \& prevention, 2023.

\bibitem{halaburda2025decentralization}
Hanna Halaburda and Daniel Obermeier.
\newblock Decentralization and the {{Law}} of the {{Jungle}}: {{An Empirical Investigation}} of {{Ethereum}}'s {{Market Mechanism}}, 2025.

\bibitem{immunebytes2021cream}
{Immunebytes}.
\newblock Cream finance exploit, oct 27, 2021 – detailed analysis.
\newblock \url{https://immunebytes.com/blog/cream-finance-exploit-oct-27-2021-detailed-analysis/}, 2021.
\newblock Accessed April 18, 2025.

\bibitem{immunefi2022beanstalk}
{Immunefi}.
\newblock Hack analysis: Beanstalk governance attack, april 2022.
\newblock \url{https://medium.com/immunefi/hack-analysis-cream-finance-oct-2021-fc222d913fc5}, 2022.
\newblock Accessed April 18, 2025.

\bibitem{nadcab2024}
Nadcab Labs.
\newblock Strategies to avoid flash loan attacks in dexs, 2024.

\bibitem{frontiers2025}
S.~Lustenberger et~al.
\newblock Dao as digital governance tool for collaborative housing.
\newblock {\em Frontiers in Blockchain}, 2025.

\bibitem{mitrade2025upcx}
{Mitrade}.
\newblock Upcx breach coverage on mitrade.
\newblock \url{https://www.mitrade.com/insights/news/live-news/article-3-733402-20250401}, 2025.
\newblock Accessed April 18, 2025.

\bibitem{verichains2022beanstalk}
{Verichains}.
\newblock How to stole an election: Beanstalk hack explained.
\newblock \url{https://blog.verichains.io/p/how-to-stole-an-election-beanstalk}, 2022.
\newblock Accessed April 18, 2025.

\bibitem{wang2024enhancing}
Zayn Wang.
\newblock Enhancing ocular diseases recognition with domain adaptive framework: Leveraging domain confusion.
\newblock {\em International Journal of Machine Learning and Cybernetics}, pages 1--9, 2024.

\bibitem{wang2022emotion}
Zeen Wang.
\newblock Emotion {{Recognition Based}} on~{{Multi-scale Convolutional Neural Network}}.
\newblock In Ying Tan and Yuhui Shi, editors, {\em Data {{Mining}} and {{Big Data}}}, pages 152--164, Singapore, 2022. Springer Nature.

\bibitem{acm2024}
Zhi Wang et~al.
\newblock Real-time disruption of non-price flash loan attacks in defi.
\newblock {\em Proceedings of the ACM Conference}, 2024.

\bibitem{ssrn2025}
David Yermack et~al.
\newblock A review of dao governance: Recent liturature and emerging trends.
\newblock {\em European Corporate Governance Institute - Finance Working Paper No. 1044/2025}, 2024.

\bibitem{arxiv2024}
Yixin Zhou and Qian Li.
\newblock Protecting defi platforms against non-price flash loan attacks.
\newblock {\em arXiv preprint arXiv:2503.01944}, 2024.

\end{thebibliography}

\end{document}